\documentclass{PoS}

\title{Extracting the strangeness freeze-out temperature from net-Kaon data at RHIC}

\ShortTitle{Net-Kaon Fluctuations}

\author{Rene Bellwied\\
        Department of Physics, University of Houston, Houston, TX, USA  77204}      
\author{\speaker{Jacquelyn Noronha-Hostler}\\
       Department of Physics and Astronomy, Rutgers University, Piscataway, NJ USA 08854\\
        E-mail: \email{jacquelyn.noronhahostler@rutgers.edu}}
\author{Paolo Parotto\\
        Department of Physics, University of Houston, Houston, TX, USA  77204}
\author{Israel Portillo Vazquez\\
        Department of Physics, University of Houston, Houston, TX, USA  77204}        
\author{Claudia Ratti\\
        Department of Physics, University of Houston, Houston, TX, USA  77204}
\author{Jamie Stafford\\
        Department of Physics, University of Houston, Houston, TX, USA  77204}

\abstract{Using the moments of the  net-kaon distribution calculated within a state of-the-art hadron resonance gas model compared to experimental data from STAR's Beam Energy Scan, we find that the extracted strange freeze-out temperature is incompatible with the light one extracted from net-proton and net-charge fluctuations.  Additionally predictions for net-$Lambda$ fluctuations are made that also appear to be consistent with a higher freeze-out temperature for strange particles.  This strangeness freeze-out temperature is roughly $10-15$ MeV higher than the corresponding light freeze-out temperature.  We also discuss cross-susceptibilities using different identified particles, which may be a further test of this two freeze-out temperature picture.  Finally, we lay out the necessary updates needed in relativistic hydrodynamic models to take into account for this two freeze-out temperature scenario and present preliminary results of $\Lambda$ spectra at RHIC for AuAu $\sqrt{s_{NN}}=200$ GeV collisions that indicate a higher freeze-out temperature is preferred. }

\FullConference{Corfu Summer Institute 2018 "The Critical Point and Onset of Deconfinement Conference"\\
		(CORFU2018)\\
		24 September - 28 September, 2018\\
		Corfu, Greece}

\begin{document}

\section{Introduction}

Relativistic heavy ion collisions have been used to probe the phase transition from a Quark Gluon Plasma into a hadron resonance gas for over 15 years now \cite{Adcox:2004mh,Back:2004je,Arsene:2004fa,Adams:2005dq}.  While early estimates for this phase transition initially assumed it was either first or second order \cite{Engels:1980ty,McLerran:1981pb,Boyd:1996bx,Huovinen:2001cy}, state-of-the-art Lattice Quantum Chromodynamic (QCD) calculations eventually proved that it was a cross-over \cite{Aoki:2006we,Borsanyi:2010cj,Borsanyi:2010bp}.  In a cross-over phase transition there is not a single  critical temperature wherein the phase transition occurs and all degrees of freedom switch between phases.  Rather, in a cross-over it is not possible to uniquely define (although in the case of QCD possibly many exist) characteristic temperatures, $T_c$, and it depends on what potential order parameter you observe.  Early Lattice QCD calculations focused on the chiral condensate, Polyakov loop, and strangeness susceptibility \cite{Aoki:2006br} and found a wide range in characteristic temperatures.    More recent Lattice QCD calculations have shown that there is approximately a 20 MeV different in the hadronization temperatures between light and strange particles \cite{Bellwied:2013cta}.  

On the experimental side of things, there has long been a tension between light and strange hadrons.  After the Quark Gluon Plasma is created in the laboratory, it expands and cools eventually reaching its hadronization temperature.  At this point the hadrons are still strongly interacting for a short period of time but quickly they reach thermal equilibrium \cite{BraunMunzinger:2003zz,NoronhaHostler:2007jf,NoronhaHostler:2009cf,Beitel:2016ghw} and "freeze-out" at a certain temperature and baryon chemical potential $\left\{T^{ch},\mu_B^{ch}\right\}$. At the point of freeze-out the  ratios of hadrons should be fixed such that one can make direct comparisons to a grancanonical ensemble (a hadron resonance gas model is used) in order to extract $\left\{T^{ch},\mu_B^{ch}\right\}$ \cite{Cleymans:1998yb,Andronic:2008gu,Manninen:2008mg,Abelev:2009bw,Aggarwal:2010pj}.  While these thermal fits have been quite successful there is still a large tension between the preferred lower freeze-out temperature of protons and pions and the higher preferred temperature of strange particles, which was dubbed the $p/\pi$ puzzle \cite{Floris:2014pta}. A number of solutions were proposed such as the inclusion of missing resonances \cite{Noronha-Hostler:2014usa,Noronha-Hostler:2014aia,Bazavov:2014xya} and finite state reactions \cite{Becattini:2016xct,Rapp:2000gy,Rapp:2001bb,Rapp:2002fc,Steinheimer:2012rd,Becattini:2012xb} wherein thermal equilibrium is never reached within the fireball. While missing resonances have now been confirmed using Lattice QCD studies of partial pressures \cite{Alba:2017mqu} the question of the unresolved tension between light and strange freeze-out temperatures still remains.

Thermal fits only utilize the mean values of particle yields, however, it is well-established in the community that event-by-event fluctuations occur that leads to fluctuations in conserved charges. Experimentally, one can measure these fluctuations by reconstructing the distribution of hadronic particles that serve as a proxy for a conserved charge.  A good example of this is that net-charge includes protons, pions, and kaons in order to reconstruct the moments of this distribution.  While heavier particles such as strange baryons certainly have a charge, they are produced significantly less due to their heavier mass so their contributions would be small.  Baryon conservation is determined using net-protons since we do not have the detector capabilities to determine neutron yields (and theoretical calculations then take isospin randomization into account) and recent results from STAR use net-kaons as a proxy for strangeness  \cite{Adamczyk:2017wsl}.  In \cite{Alba:2015iva} it was found that direct comparisons between a hadron resonance gas model and net-Q and net-p can be used to determine the freeze-out temperature of light hadrons. The temperature in \cite{Alba:2015iva} was a lower temperature than that indicated from thermal fits.  Lattice QCD has also made direct comparisons between susceptibilities of the pressure to net-Q and net-p \cite{Borsanyi:2014ewa} and these results are consistent with those found in \cite{Alba:2015iva}. In \cite{Noronha-Hostler:2016rpd} a new technique was used to isolate the moments of the net-kaon distribution directly from Lattice QCD but these calculations have not yet been performed for all the beam energies at RHIC.

In this talk based on \cite{Bellwied:2018tkc}  net-Kaon fluctuations were used to extract the strange freeze-out temperature $\left\{T^{ch},\mu_B^{ch}\right\}_s$ wherein we find that it is incompatible with the light freeze-out temperature (roughly $\Delta T\sim 10-15 $ MeV at top RHIC energies.  However, at lower beam energies signs of a convergence in the freeze-out temperatures is displayed.  We also show preliminary results for exploring effects of the freeze-out temperature in hydrodynamics.

\section{net-Kaon fluctuations}

Within the hadron resonance gas model the pressure can be determined by summing over quantum numbers of all known hadrons such that:
\begin{equation}
p(T,\mu_B,\mu_Q,\mu_S) =  \sum_{i \in {HRG}} (-1)^{B_i+1} \frac{d_iT}{(2\pi)^3} \int {\rm d}^3\vec{p} \,\,\ln\left[1+(-1)^{B_i+1} \exp({-(\sqrt{\vec{p}^{\,2}+m_i^2}-B_i\mu_B-S_i\mu_S-Q_i\mu_Q)/T})\right] .
\label{equ:pressureHRG}
\end{equation}
Then one can take derivatives of the pressure in respect to a certain conserved charge, which are known as susceptibilities.  In this case we are only concerned about net-strangeness and specifically that of net-Kaons so only charged kaons and resonances that decay into them are considered:
\begin{eqnarray}\label{eqn:chis}
\chi_n^{net-K}&=&\!\!\!\sum_{i \in HRG} \!\!\!\! \frac{\left(Pr_{i\rightarrow {net-K}}\right)^n}{T^{3-(n-1)}} \frac{S_i^{1-n}d_i}{4\pi^2}\frac{\partial^{n-1}}{\partial \mu_S^{n-1}}\left\{ \int_{-0.5}^{0.5} \!\!\!\!\!\! {\rm d}y \int_{0.2}^{1.6} \!\!\!\!\!\! {\rm d}p_T\right. \times \nonumber\\
&\times& \left.\frac{p_T\sqrt{p_T^2+m_i^2}{\rm Cosh}[y]}{(-1)^{B_i+1}+\exp({({\rm Cosh}[y]\sqrt{p_T^{\,2}+m_i^2}-(B_i\mu_B+S_i\mu_S+Q_i\mu_Q))/T})}  \right\}.
\end{eqnarray}
One can then make direct comparisons between the susceptibilities of pressure and experimental moments of the net-Kaon distribution via: $\chi_1^K/\chi_2^K=(M/\sigma^2)_K$.  

However, in order to extract both $\left\{T^{ch},\mu_B^{ch}\right\}_s$ we require two observables since there are two unknowns. Thus the results for $(M/\sigma^2)_K$ are insufficient to do this.  For the light sector previous results were able to constrain $\left\{T^{ch},\mu_B^{ch}\right\}_l$ using both net-p and net-Q.  Here we exploit isentrope trajectories.

In  \cite{Alba:2015iva} the light freeze-out $\left\{T^{ch},\mu_B^{ch}\right\}_l$ was used to determine isentrope trajectories \cite{Gunther:2016vcp} from the taylor reconstructed Lattice QCD results at finite $\mu_B$ where the entropy to baryon number was held constant i.e $S/N_B=const$.  While $S/N_B$ is not exactly conserved due to entropy production from viscosity, since $\eta/s$ and $\zeta/s$ are quite small and we are only considering here a small section of the phase diagram we have estimated the effect to be extremely small ($\sim1\%$).  Thus using both the isentrope trajectories and the net-Kaon data we are able to extract $\left\{T^{ch},\mu_B^{ch}\right\}_s$.

\begin{figure*}[t]
\centering
\includegraphics[width=\linewidth]{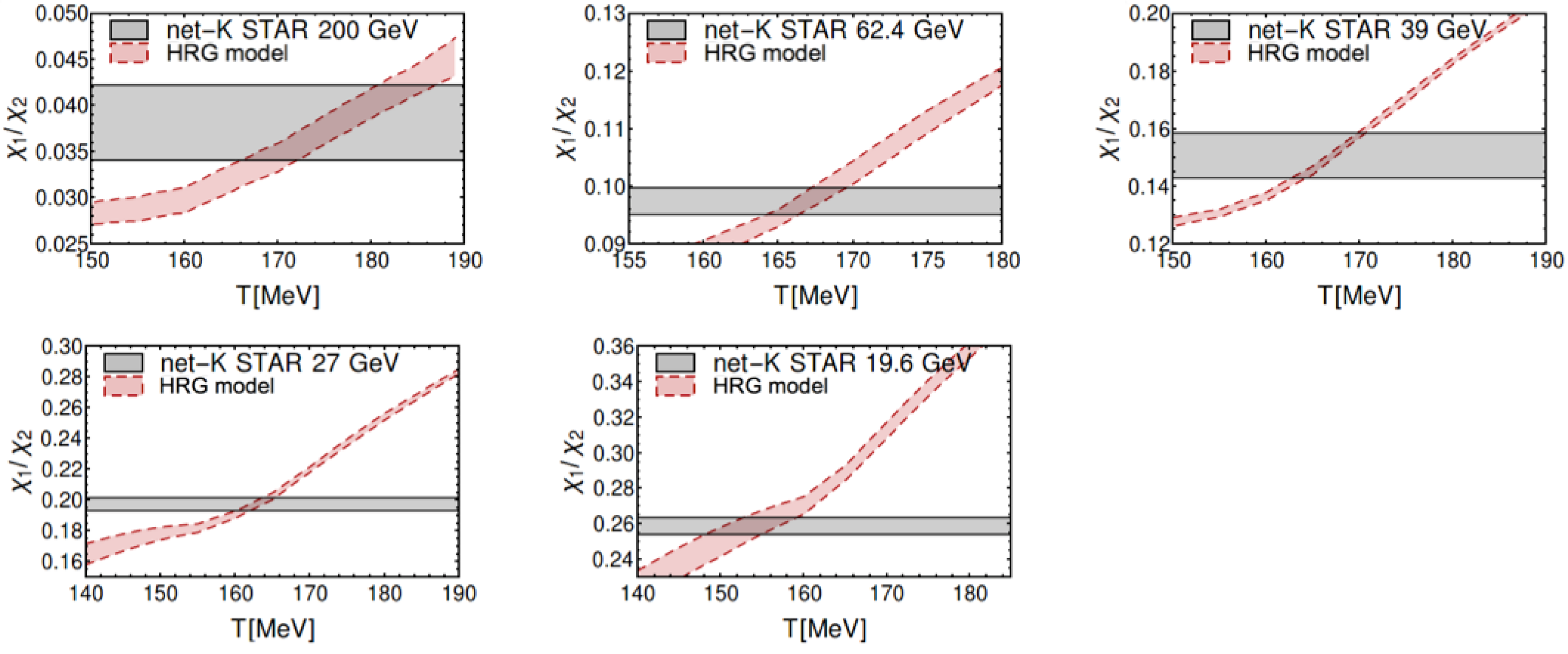} 
\caption{Results for $\chi_1^K/\chi_2^K$ calculated in the HRG model along the Lattice QCD  isentropic trajectories  (pink, dashed band) compared to  $(M/\sigma^2)_K$ data from \cite{Adamczyk:2017wsl} (gray, full band) across the Beam Energy Scan at STAR.} \label{fig:isen}
\end{figure*}

In Fig.\ \ref{fig:isen} the isentrope trajectories are used to calculate $\chi_1^K/\chi_2^K$ and compare directly to STAR data at each center of mass energy $\sqrt{s_{NN}}$. The overlapping region between the two curves is the allowable range in strange freeze-out temperatures. While at $\sqrt{s_{NN}}=200$ GeV the allowable temperature range from these measurements reaches quite high, those highest temperatures are unlikely due to their incompatibility with Lattice QCD strange particle hadronization temperature estimates \cite{Bellwied:2013cta}. 

\begin{figure}[h]
\centering
\includegraphics[width=0.5\linewidth]{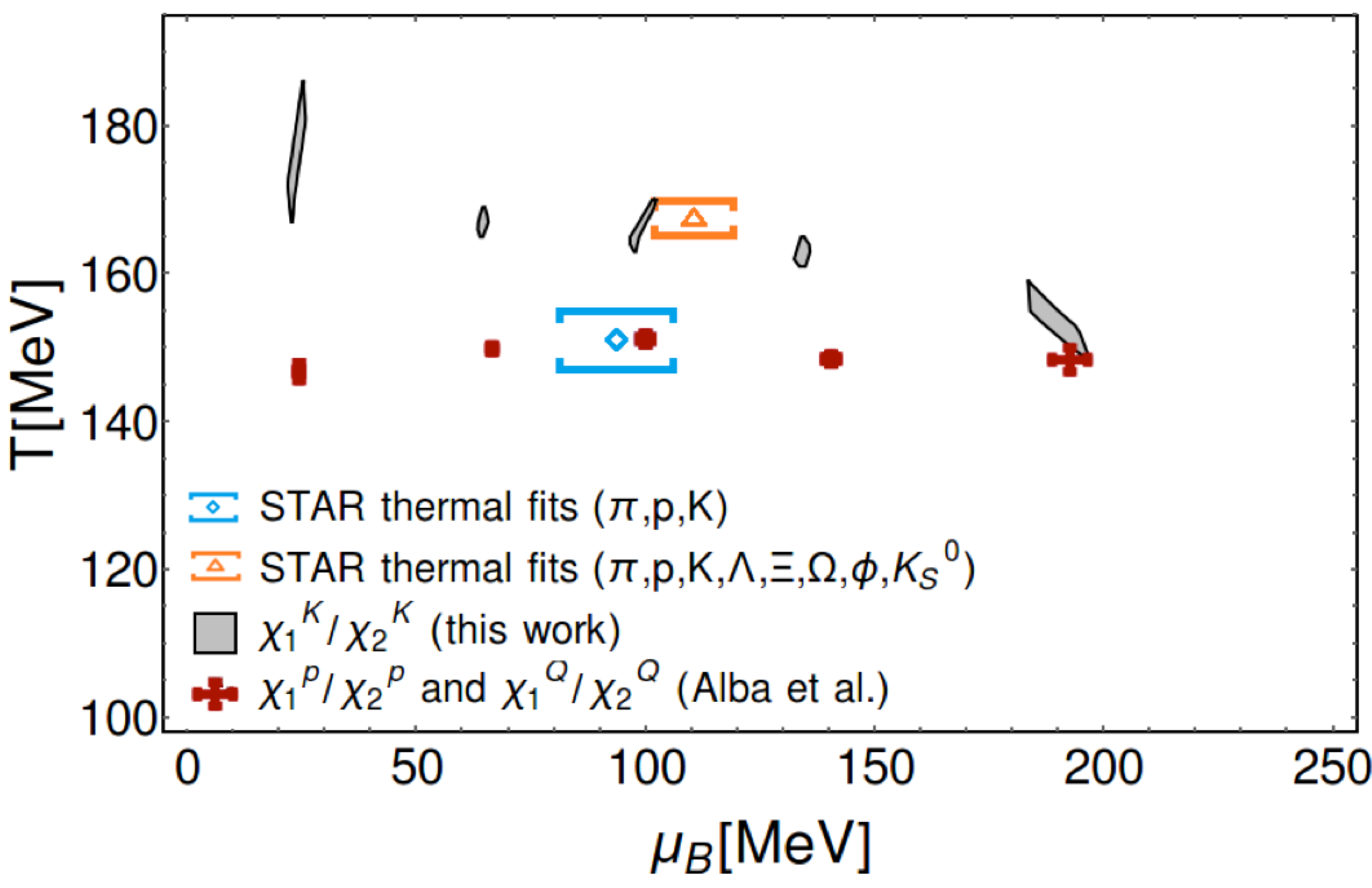} 
\caption{Freeze-out parameters across the highest five energies from the Beam Energy Scan. The red points were obtained from the combined fit of $\chi_1^p/\chi_2^p$ and $\chi_1^Q/\chi_2^Q$ \cite{Alba:2014eba}, while the gray bands are obtained from the fit of $\chi_1^K/\chi_2^K$ in this work. Also shown are the freeze-out parameters obtained by the STAR collaboration at $\sqrt{s}=39$ GeV \cite{Adamczyk:2017iwn} from thermal fits to all measured ground-state yields (orange triangle) and only to protons, pions and kaons (blue diamond-shaped symbol). Taken from \cite{Bellwied:2018tkc}.} \label{fig:Tmub}
\end{figure}

Then the corresponding $\left\{T^{ch},\mu_B^{ch}\right\}_l$ from \cite{Alba:2014eba} and $\left\{T^{ch},\mu_B^{ch}\right\}_s$ from \cite{Bellwied:2018tkc}  are shown in Fig.\ \ref{fig:Tmub} where a distinct discrepancy exists between the two temperatures across beam energies.  Only at the lowest beam energies do we see an hint of a convergence. Interesting enough, thermal fit results from  \cite{Adamczyk:2017iwn} with and without strange baryons at $\sqrt{s}=39$ GeV appear to coincide almost perfectly with our corresponding light and strange freeze-out temperatures.  This indicates that further explorations between thermal fits with 2 freeze-out temperatures and fluctuations of conserved charges may resolve previous discrepancies between the two. 

Independently another group found similar results in \cite{Bluhm:2018aei} using net-Kaon fluctuations and the yields of strange baryons.  There they also considered probabilistic contributions to the net-kaon fluctuations, which increased the strangeness freeze-out temperature even further ($\sim 5\%$).

\begin{figure*}[t]
\centering
\begin{tabular}{c c}
\includegraphics[width=0.5\linewidth]{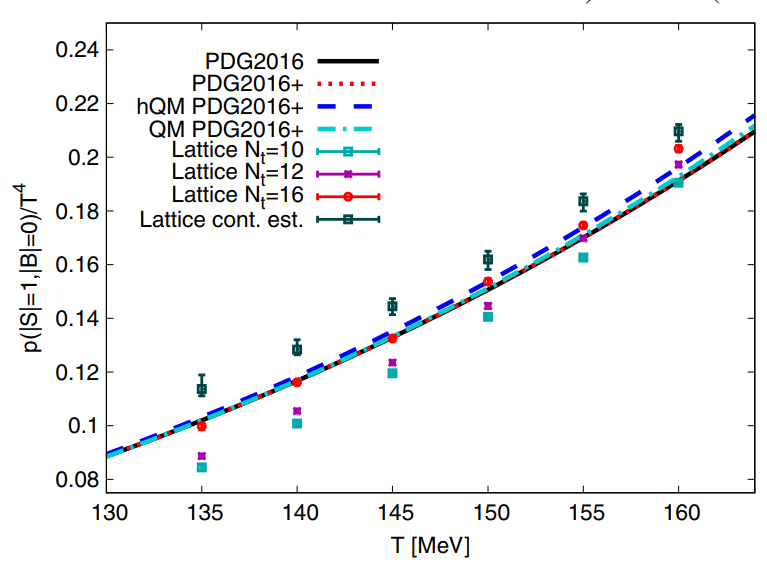}  & \includegraphics[width=0.5\linewidth]{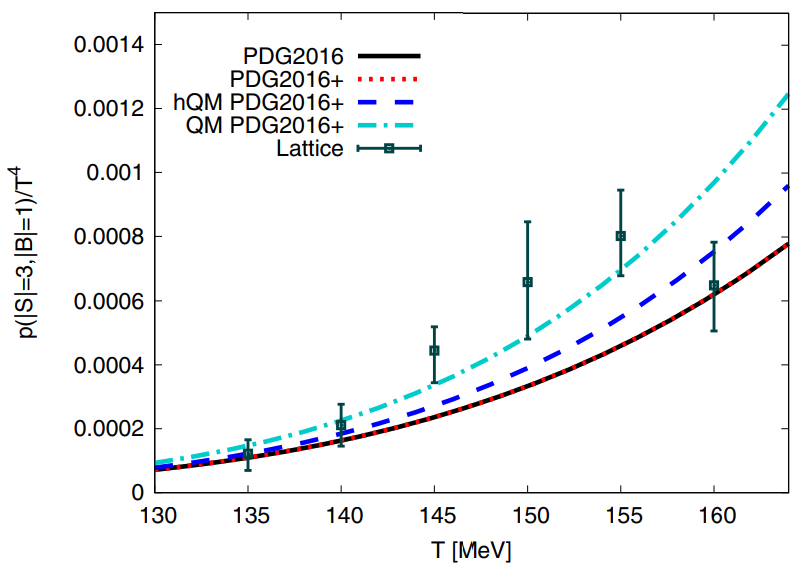}
\end{tabular}
\caption{Comparison of the pressure pressure contribution to strange mesons  (left) and S=3 baryons (right) between Lattice QCD results \cite{Alba:2017mqu} and hadron resonance gas models with varying particle lists. Taken from \cite{Alba:2017mqu}. } \label{fig:kmiss}
\end{figure*}

The one caveat to this work is the possibility of missing strange mesonic resonances as compared to particle pressure calculated from Lattice QCD in \cite{Alba:2017mqu}. Multiple different particle lists were compared to the Lattice QCD data and all under-predicted the strange meson partial pressure, as shown in Fig.\ \ref{fig:kmiss}.  However, on the Lattice QCD end it was not possible to yet extrapolate into the continuum limit so it is unclear if this may still be a truncation error or if it is an actual physics effect.  One should note, however, that the different particle lists considered in \cite{Alba:2017mqu} did have large effects in the strange baryon regime as shown on the right in Fig.\ \ref{fig:kmiss}.

\section{net-Lambda predictions}

\begin{figure}[h]
\centering
\begin{tabular}{c c}
\includegraphics[width=.5\linewidth]{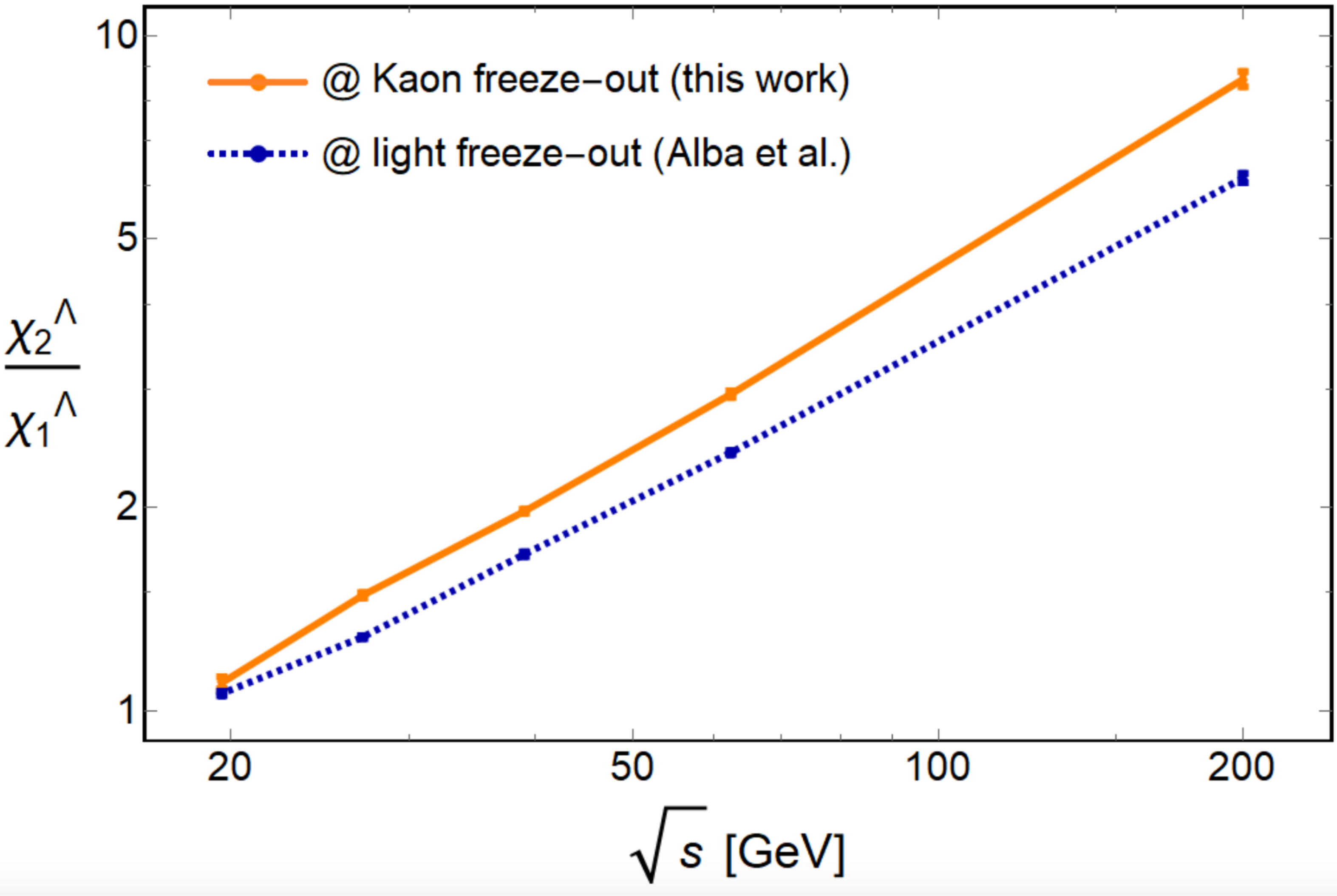} &
\includegraphics[width=.5\linewidth]{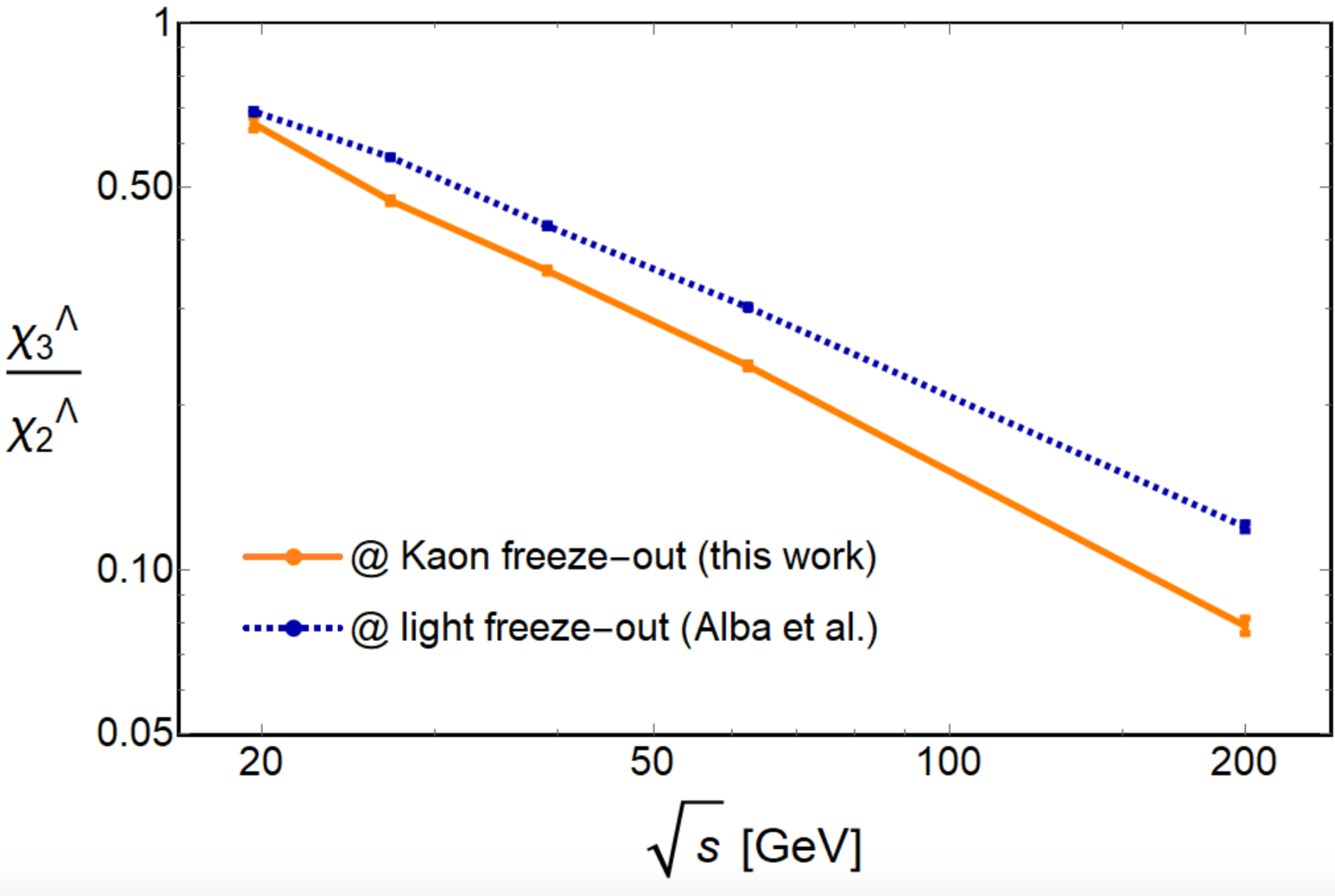}
\end{tabular}
\caption{Upper panel: $\chi_2^{\Lambda}/\chi_1^{\Lambda}$ as a function of $\sqrt{s}$. Lower panel: $\chi_3^{\Lambda}/\chi_2^{\Lambda}$ as a function of $\sqrt{s}$. In both panels, the orange points joined by a full line are calculated at the values of $T_f$ and $\mu_{Bf}$ extracted from the fit of $\chi_1^K/\chi_2^K$, while the blue points joined by a dashed line are calculated at the values of $T_f$ and $\mu_{Bf}$ extracted from the combined fit of $\chi_1^p/\chi_2^p$ and $\chi_1^Q/\chi_2^Q$ in Ref. \cite{Alba:2014eba}. Taken from \cite{Bellwied:2018tkc}.} \label{Lambda}
\end{figure}

While we can extract the strange temperature from already existing experimental data, a much stronger piece of evidence for the flavor hierarchy is when predictions are made and then later confirmed.  In \cite{Bellwied:2018tkc} we predicted the $(M/\sigma^2)_\Lambda$ of net-Lambda fluctuations using both the light freeze-out point $\left\{T^{ch},\mu_B^{ch}\right\}_l$ from \cite{Alba:2014eba} and the strangeness freeze-out $\left\{T^{ch},\mu_B^{ch}\right\}_s$ extracted from net-K.  In Fig.\ \ref{Lambda} one can see that there is a clear difference between two temperatures wherein a higher strange freeze-out temperature indicates a wider distribution in net-strangeness fluctuations.  Preliminary results of net-$\Lambda$ fluctuations were shown at Quark Matter 2018, which are compatible  with the higher strangeness freeze-out temperature, not the light freeze-out temperature. UrQMD results were found to be incompatible with the STAR net-$\Lambda$ $(M/\sigma^2)_\Lambda$, especially at top RHIC energies. 

\section{Off-diagnoal predictions}

\begin{figure*}[t]
\centering
\includegraphics[width=0.5\linewidth]{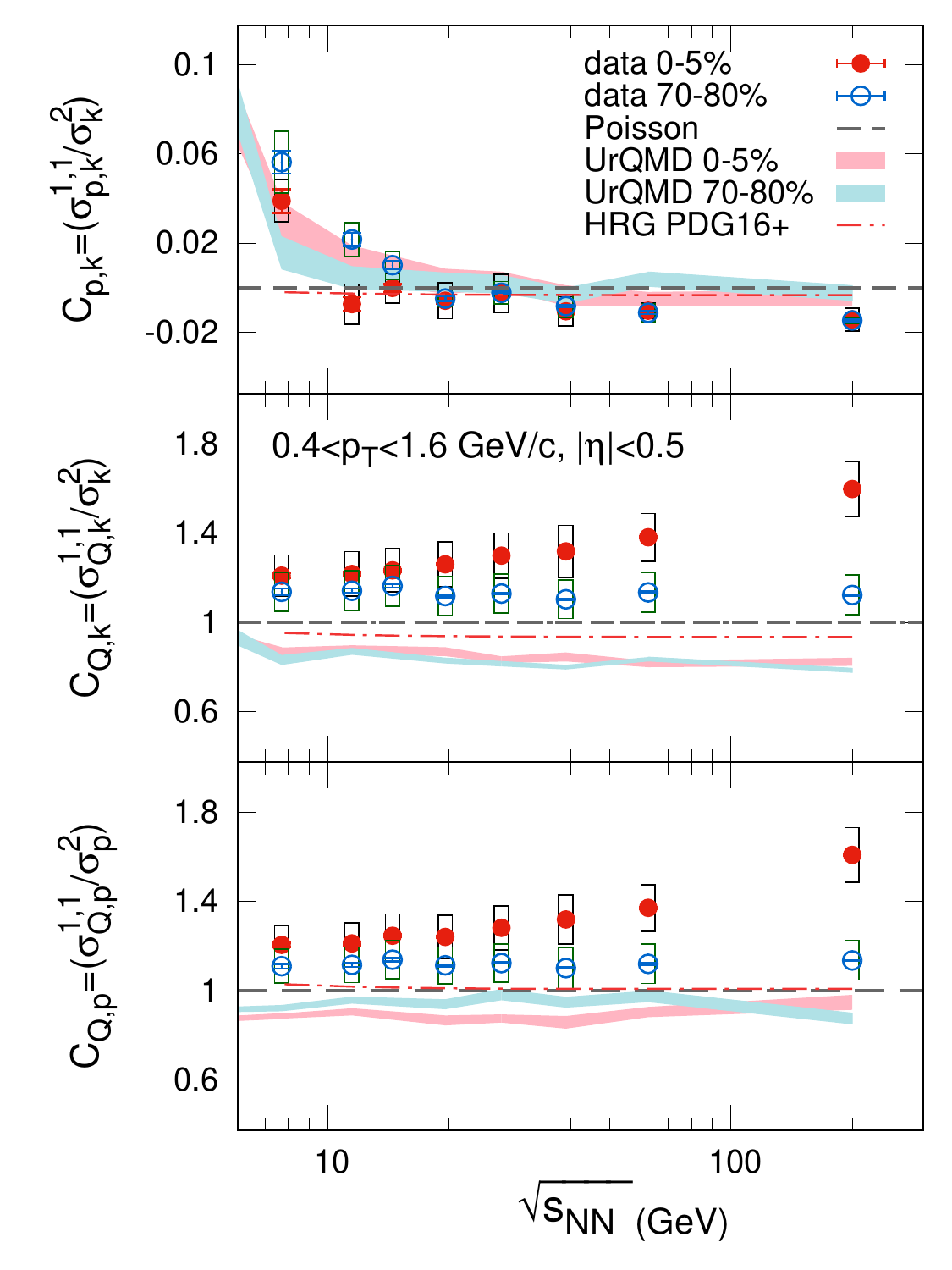} 
\caption{Predictions made from PDG16+ for cross-cumulants using the light freeze-out temperature compared to STAR data \cite{Adam:2019xmk}. Taken with permission of the authors from Fig 6 in \cite{Adam:2019xmk}.  } \label{fig:cross}
\end{figure*}

Another possibility to explore the interplay between light and strange particles is to use cross-cumulants between different identified particles that represent different conserved charges.  To do so we generalize our susceptibilities such that:
\begin{equation}\label{eqn:chis}
\chi_{11}^{\Delta A\Delta B}=\sum_i^{N_{HRG}}\left(Pr_{i\rightarrow A^{\pm}}  X_i\right)\left(Pr_{i\rightarrow B^{\pm}}  Y_i\right) \frac{d_i}{4\pi^2} \int_{-0.5}^{0.5} {\rm d}\eta \int_{0.2}^{1.6} {\rm d}p_T\times \frac{ p_T^2{\rm Cosh}[\eta]\exp\left({\frac{\sqrt{p_T^2{\rm Cosh}^2[\eta]+m^2_i}}{T}-\hat{\mu}}\right)}{\left[(-1)^{B_k+1}+\exp\left({\frac{\sqrt{p_T^2{\rm Cosh}^2[\eta]+m^2_i}}{T}-\hat{\mu}}\right)\right]^2} 
\end{equation}
where B is represented by protons, S by kaons, and Q by protons, pions, and kaons. Then ratios of the susceptibilities are calculated such that 
\begin{equation}
C_{A,B}=\frac{\chi_{11}^{\Delta A\Delta B}}{\sigma^B_2}
\end{equation}
 Using the light freeze-out temperature in \cite{Alba:2014eba} we made predictions for the PDG16+ for a recent STAR paper \cite{Adam:2019xmk}. Our predictions are shown in Fig.\ \ref{fig:cross}.  From the comparisons to data while the HRG results appear to be in the right order of magnitude we are unable to capture the temperature dependence at the highest RHIC energies in the $0-5\%$ centrality class so this maybe leave room to explore different freeze-out temperatures but we are still trying to understand these deviations.

\section{2 Freeze-out temperatures in Hydrodynamics}

While the evidence of two separate freeze-out temperatures is becoming more solid, the ability to model this consistently in relativistic viscous hydrodynamical models is still beyond our current capabilities.  In order to achieve this goal significant upgrades must be made to hydrodynamic codes that are outlined here:
\begin{itemize}
\item New Equation of State with finite baryon number and strangeness
\item Strangeness and baryon number conservation 
\item Strangeness and baryon number diffusion
\item Initial conditions with a strange and baryon number distribution
\item Hadronization that considers charge conservation
\end{itemize}

A number of groups are currently working towards these goal.  For instance, within the last few months two papers have recently come out about the development of an equation of state with 3 conserved charges \cite{Noronha-Hostler:2019ayj,Monnai:2019hkn}. In both cases only a small region in the cross-over transition is covered due to the limitation in the number of calculated susceptibilities on the lattice.  Unlike in the case where an equation of state is constructed in just the $\{T,\mu_B\}$ plane where only a single critical point may exist \cite{Critelli:2017oub,Parotto:2018pwx}, if one considers all 3 conserved charges it may be a critical line or even a plane.   Thus, reconstructing an equation of state with multiple conserved charges combined with criticality is a significant challenge that still remains. 

Additionally, it is unlikely that an equation of state that switches the strange particles to the hadron gas phase at a higher temperature than light particles will be produced in the near future.  This would require new Lattice QCD results  with just 2 flavors and we are not aware of a Lattice QCD collaboration who is running these results at this time. But because of how well the hadron resonance gas matches Lattice QCD results precisely in this regime, it is unlikely that significant differences would be seen from the equation of state alone. 

Conserved charges (both in terms of the number conservation and transport coefficients i.e. baryon and strangeness diffusion) are currently be implemented into a number of relativistic hydrodynamics codes \cite{Monnai:2016kud,Greif:2017byw,Denicol:2018wdp} and it is only a matter of time before this is widespread amongst the community. The calculations of baryon, strangeness, and electric charge diffusion transport coefficients are also being explored by a wide variety of groups \cite{Rougemont:2015ona,Greif:2016skc,Rougemont:2017tlu,Greif:2017byw} and of particular interest on differences in the $T_c$ of these different transport coefficients \cite{Rougemont:2017tlu}. 

What is not quite as trivial is the initialization of conserved charges which has mostly be done either from a parton or string picture \cite{Werner:1993uh,Pierog:2013ria,Petersen:2008dd,Shen:2017bsr} but current efforts are underway to initialize conserved charges within a color glass condensate as well (from gluon splitting) \cite{Martinez:2018tuf}. 

Finally, one of the biggest unsolved problems when it comes to two separate freeze-out temperatures is the proper manner in which one should hadronize strange and light particles separately.  A clear consensus has not yet been achieved on this point.

\begin{figure*}[t]
\centering
\includegraphics[width=0.5\linewidth]{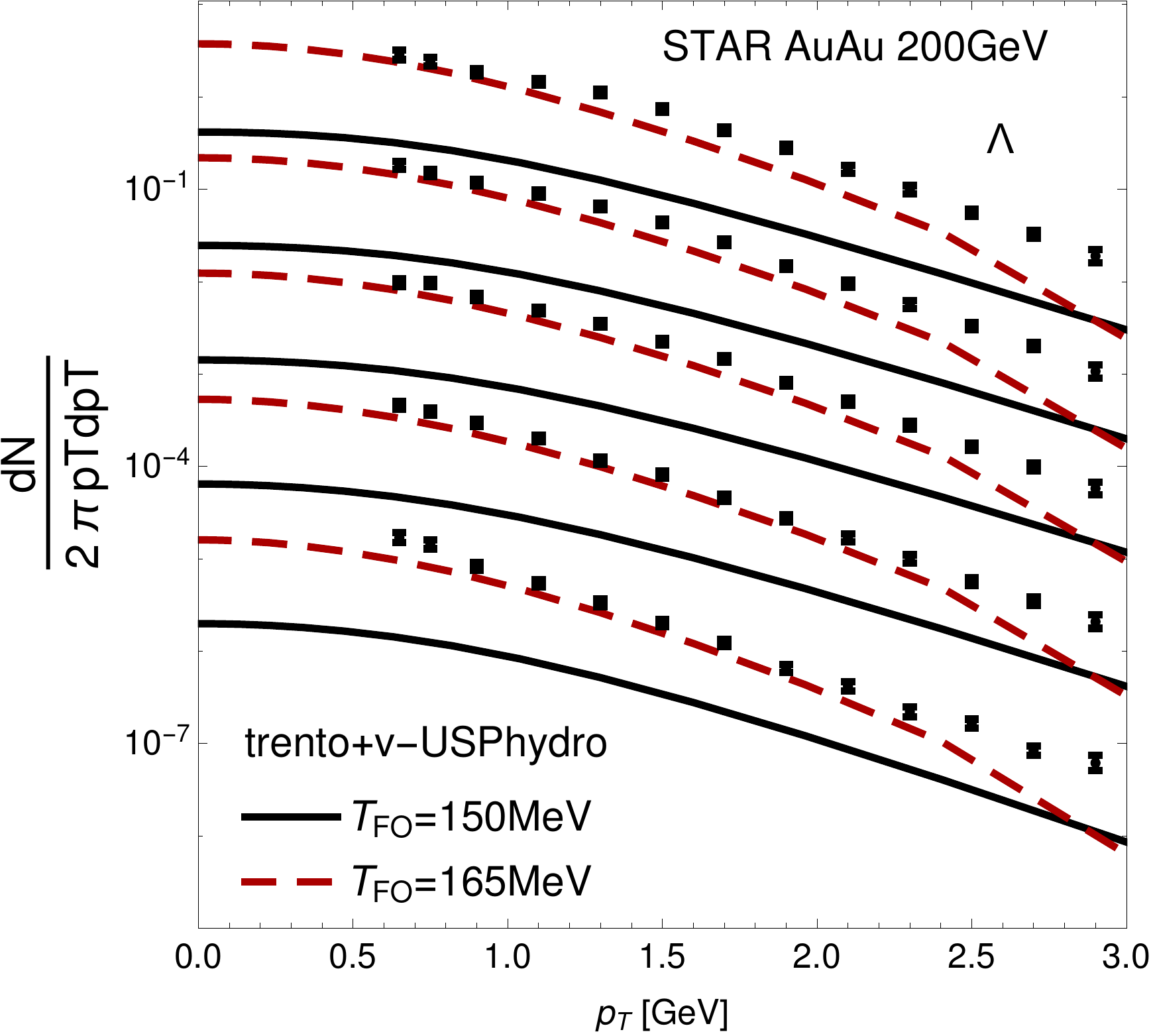} 
\caption{Trento+v-USPhydro calculations at two different freeze-out temperatures: $T_{FO}=150$ MeV and  $T_{FO}=165$ MeV  compared to experimental spectra data from STAR \cite{Estienne:2004di} for AuAu $\sqrt{s_{NN}}=200$ GeV} \label{fig:2FO}
\end{figure*}

With all these caveats in mind, we wanted to first determine if the strange sector had a sensitivity to its freeze-out temperature or not. In order to study this we used trento+v-USPhydro \cite{Noronha-Hostler:2013gga,Noronha-Hostler:2014dqa} that includes particle decays from the PDG16+ \cite{Alba:2017mqu}.  Previous results of pions, protons, and kaons have shown good fits to experimental data using $T_{FO}=150$ MeV \cite{Alba:2017hhe}.  However, as see in Fig.\ \ref{fig:2FO}, these same results for the spectra fail for $\Lambda$'s and significantly under-predict the spectra.  However, if we use $T_{FO}=165$ MeV the fits look reasonable compared to experimental data. Therefore we must conclude that this effort to implement two separate freeze-out temperatures for light and strange holds promise. 

\section{Conclusions}
In this proceedings we extracted the strange freeze-out temperature using net-kaon fluctuations data from STAR.  We find that these temperatures are roughly 20 MeV higher than the light temperatures at the highest RHIC beam energies and are, therefore, incompatible with a single freeze-out temperature picture.  Predictions made for net-$\Lambda$ fluctuations are also consistent with a flavor hierarchy and we are checking cross-correlations as well to see if they can further constrain the flavor hierarchy  scenario.  

Additionally, we detailed the necessary steps needed to accurately model the flavor hierarchy dynamically within relativistic hydrodynamic models.  As an exploratory study we tested two different strange switching temperatures in hydrodynamics for AuAu $\sqrt{s_{NN}}=200$ GeV at RHIC  and found that only $T=165$ MeV was compatible with the spectra data whereas $T=150$ MeV, which has been previously shown to reproduce the light hadron spectra well \cite{Alba:2017hhe}.  While more systematic studies are required in the future this proves to be a promising path to explore. 

\section*{Acknowledgements}
This  material  is  based  upon  work  supported  by  the National  Science  Foundation  under  grants  no.    PHY-1654219 and OAC-1531814 and by the U.S. Department of  Energy,  Office  of  Science,  Office  of  Nuclear  Physics, within the framework of the Beam Energy Scan Theory (BEST) Topical Collaboration.  We also acknowledge the support from the Center of Advanced Computing and Data Systems at the University of Houston. The work of R. B. is supported through DOE grant DEFG02-07ER41521. J.N.H. acknowledges support from the US-DOE Nuclear Science Grant No. DE-SC0019175 and the Office of Advanced Research Computing (OARC) at
Rutgers, The State University of New Jersey for providing
access to the Amarel cluster and associated research
computing resources that have contributed to the results
reported here.

\bibliographystyle{JHEP}
\bibliography{biblio}

\end{document}